**Surface engineering for ultrathin metal anodes enabling high-performance Zn-ion batteries**

*Ziyi Hu, Linming Zhou, Dechao Meng, Liyan Zhao, Yihua Li, Yuhui Huang\*, Yongjun Wu\*,*

*Shikuan Yang, Linsen Li\*, Zijian Hong\**

Z. Hu, L. Zhou, Y. Li, Prof. Y. Huang, Prof. Y. Wu, Prof. Z. Hong

Cyrus Tang Center for Sensor Materials and Applications, State Key Laboratory of Silicon

Materials, School of Materials Science and Engineering, Zhejiang University, Hangzhou 310027,

China

E-mail: yongjunwu@zju.edu.cn (Y. W.), huangyuhui@zju.edu.cn (Y. H.),

hongzijian100@zju.edu.cn (Z. H.)

L. Zhao, Prof. S. Yang

Lab of Composite Materials, School of Materials Science and Engineering,

Zhejiang University, Hangzhou 310027, China

D. Meng, Prof. L. Li

Department of Chemical Engineering, Shanghai Jiao Tong University, Shanghai 200240, China

E-mail: linsenli@sjtu.edu.cn

**Abstract:** Zn metal battery has been considered a promising alternative energy storage technology in renewable energy storage and grid storage. It is well-known that the surface orientation of a Zn metal anode is vital to the reversibility of a Zn metal battery. Herein, the (101)-oriented thin Zn metal anode (down to 2 μm) is electrodeposited on a Cu surface by adding dimethyl sulfoxide (DMSO) electrolyte additive in $ZnSO_4$ aqueous solution. Scanning electron microscope (SEM) observation indicates the formation of flat terrace-like compact (101)-oriented surfaces. Insitu optical observation confirms that the (101)-oriented surfaces can be reversibly plated and stripped. DFT calculations reveal two mechanisms for the nucleation and growth of the Zn-(101) surface: (1) formation of Zn(101)//Cu(001) could lower the interface energy as compared to Zn(002)//Cu(001); (2) large reconstruction of the Zn (101) surface with DMSO and H2O absorption. Raman, XPS, and ToF-SIMS characterizations indicate that adding DMSO in ZnCl2 could facilitate the formation of ZnO-based SEI on Zn metal surface, while OH- and S-based SEI can be obtained with DMSO in $ZnSO_4$. The electrochemical testings are performed, which demonstrates a higher cyclability for the (101)-oriented Zn in the half cell as well as a lower charge transfer barrier with respect to the (002)-dominated surface of the same electrode thickness. $Zn\|V_2O_5$ full cells are further assembled, showing better capacity retention for the (101)-Zn as compared to the (002)-Zn with the same thickness (5 μm, 3 μm, and 2 μm). We hope this study to spur further interest in the control of Zn metal surface crystallographic orientation towards ultrathin Zn metal anodes.

## 1. Introduction

Zn-ion batteries with low cost, high safety, high volumetric capacity, and low toxicity have been regarded as one of the most promising alternative energy storage technologies for applications in renewable energy storage and grid storage[1]. They can use nonflammable aqueous electrolytes, which could effectively avoid serious safety issues such as fire and explosion. Moreover, unlike chemically active lithium metal, Zn metal is more stable and easier to store, transport, and recycle, making it an ideal anode for a Zn-ion battery. However, one challenge that hinders the commercialization of the Zn-ion battery is the poor reversibility, which is associated with the Zn dendrite growth during electrodeposition[2] as well as the corrosion of the Zn metal[3] in an acidic electrolyte that can result in $H_2$ gas generation and irreversible Zn loss[4].

Previously, it is shown that the surface orientation of a Zn metal anode is vital to the reversibility of a Zn metal battery[1e, 5]. With a hexagonal symmetry, the most stable low index surfaces for Zn are the (002) and (101) surfaces (under the Miller index convention). Zheng *et al.*

discovered that the (002) surface is more compact with lower surface energy, which could promote the stable electrodeposition via an epitaxial growth mechanism, achieving exceptional reversibility over thousands of plating/stripping cycles [1e]. Numerous methods have been applied to increase the (002) surface ratio of the Zn metal anode, including substrate regulation[6], electrolyte engineering (with additives[7] or gel-electrolytes[8]), and surface protective-layer coating[9]. Meanwhile, so far, the (101)-oriented surface of a Zn metal anode is less well understood. Investigating the (101)-oriented Zn metal surface could enable a comprehensive understanding of the thermodynamics and kinetics of the electrodeposition process, in particular, the competition between the nucleation and growth of the (002)-oriented and (101)-oriented surfaces during electrodeposition and their impact on dendrite growth[10]. One intriguing question to ask: can we engineer (101)-oriented Zn metal anode for a stable Zn metal battery?

Electrodeposition is a facile synthesis technique to obtain metal anodes, where thin metal atoms are deposited on an electrode by applying an electric current through an electrochemical cell. The surface orientation of the metal anodes can be engineered by tuning the electrodeposition parameters[11] (e.g., deposition current density, deposition time, electrolyte composition, etc.). For instance, it is demonstrated that a (110) surface-dominated lithium metal anode can be obtained by increasing the electroplating capacity, which displays fast deposition/stripping kinetics and excellent reversibility in a battery [10, 12]. In another example, it is shown that the crystallographic orientation of Zn deposits is significantly affected by the cation structures and related physicochemical properties of the ionic liquid electrolytes [13]. Moreover, by adding a small amount of electrolyte additives, the surface composition (e.g. solid electrolyte interface, SEI) and surface orientation of the electrodeposited Li and Zn metals can be altered. In particular, dimethyl sulfoxide (DMSO) with a large Guttman donor number has been widely employed as electrolyte additives in lithium-metal and lithium-air batteries, which could modify the solvation structure and composition. Recently, Cao *et al.* demonstrated that by adding DMSO to the $ZnCl_2$ aqueous electrolyte, a protective SEI formed as a result of the change in the solvation shell, yielding high reversibility for the Zn metal battery[7a]. However, the extent to which the modification in solvation structure could affect the characteristics of a Zn metal surface orientation remains unclear.

Herein, we employed a combination of experimental observations and theoretical calculations to investigate the influence of the electrolyte additive on the surface orientation of a Zn metal anode, as well as the impact of the surface orientation on the electrochemical performance of a Zn metal battery. Electrodeposition of thin Zn metal down to 2 μm is achieved on the Cu electrode, where the Zn metal surface orientation is controlled by electrolyte additive (DMSO). Without additive, the (002)-surface dominates in the electrodeposited Zn metal anode, with regional loose hexagonal crystals; while with 5% DMSO added in the 2 M ZnSO$_4$ solution, the (101)-surface is more favorable, forming a compact surface with parallel habit crystal planes. Contradictory to popular belief, electrochemical testings indicate that the (101)-oriented Zn metal exhibits superior reversibility with small polarization in a symmetrical battery. While the full battery is further assembled with thin electrodeposited Zn metal anode (2 μm, 3 μm, and 5 μm), commercially available V$_2$O$_5$ cathodes, and 2 M Zn(OTf)$_2$ electrolyte, which shows a maximum specific capacity of 359 mAh/g, and 170 mAh/g after 300 cycles, that is comparable to the commercially available 20 μm Zn metal foil. We hope this study could spur further interest in the orientation-dependent of the electrochemical performance of metal anodes.

## 2. Results and Discussion

*Surface morphological and structural characterization*. Field-emission scanning electron microscopy (SEM) is employed to characterize the surface morphology of the as-deposited Zn metal (denoted by D-Zn) surface **(Figure 1)**. The planar view SEM image of the as-deposited Zn metal surface without DMSO can be observed in **Figure 1(a)**, showing disordered sub-micron flake-like crystal planes. Meanwhile, with the addition of DMSO in the electrolyte (2.5%, 5%, and 7.5% in vol%, **Figure 1b-1d**), the D-Zn metal surface is occupied with mainly well-ordered compact crystal stacks, forming terrace-like surface structures. Among them, with a 5 % volume fraction of DMSO in the electrolyte, the D-Zn metal surface demonstrates minimum surface unevenness (**Figure 1c**). The cross-section views of the SEM and EDS images for D-Zn anode (deposited with 5 % DMSO in an aqueous electrolyte) are shown in **Figure 1(e)**, which indicates the formation of a uniform interface with an average of ~6.0 μm thick Zn metal on top of a 20 μm Cu substrate. The low magnification cross-section view of the electrodeposited layer is further given in **Figure S1**, confirming the uniformity of the deposited Zn metal anode. It should be noted here that the deviation with the theoretical deposition thickness of 5.0 μm is caused by the micropores and slight surface unevenness[11]. To further quantify the ratio of different surface orientations for the D-Zn, X-ray diffraction (XRD) characterization is performed. As shown in **Figure 1(f)**, the three diffraction peaks between 35° to 45° can be indexed to the Cu (111) and (002) surfaces (JCPDF# 04-0836), as well as the (100) and (101) surfaces of Zn (JCPDF#87-0173). No impurity peaks can be seen. Notably, it can be discovered that the intensity ratio of the two Zn peaks, i.e., I(101)/I(002), increases with increasing DMSO volume fraction until reaching the maximum of 8.82 with 5 % DMSO in the electrolyte, which then decreases to 4.79 for 7.5 % DMSO addition. This indicates that the dense surface shown in **Figure 1(c)** is mainly (101)-oriented, in good agreement with the tilting of the crystal plane with respect to the *xy* surface.

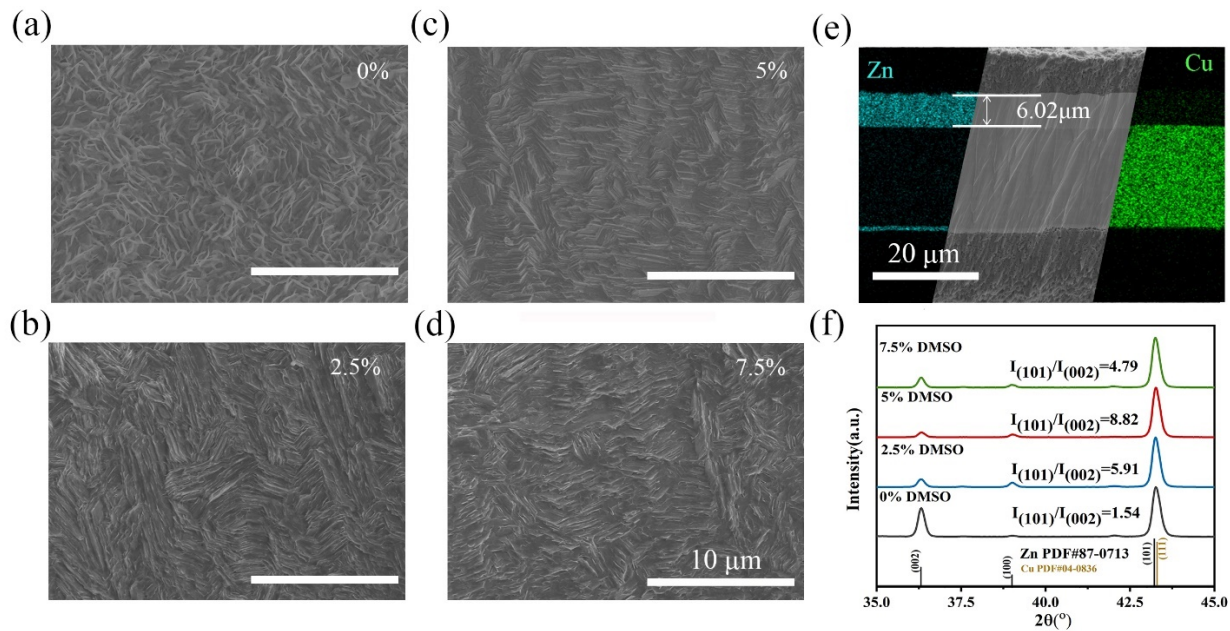

**Figure 1| Morphological and structural characterization of the as-deposited Zn metal anode.** SEM images of Electrodes through electrodeposition (a) without DMSO addition. (b) with 2.5 vol.% DMSO addition (c) with 5 vol.% DMSO addition(d) with 7.5 vol.% DMSO addition. (e) SEM and EDS elemental mapping of an ultrathin zinc electrode in cross-section view, while the line indicates the thickness of Zn. (f) XRD of the above electrodes and the peak intensity ratio of the Zn I(101)/I(002).

As a comparison, the surface morphology for the electrodeposited Zn metal with other aqueous electrolytes is given in **Figure S2**. It can be observed that an island-like loose surface formed with 1.3 M $ZnCl_2$ as the deposition electrolyte (**Figure S2a**). Whereas the addition of an 18.75% volume fraction of DMSO could lead to the formation of more compact surfaces, with the tilting of (101)-oriented surfaces (**Figure S2b**). Meanwhile, using 2M $Zn(OTf)_2$ electrolyte, the surface morphology is flatter as compared to $ZnCl_2$, as shown in **Figure S2(c)**. While adding a 5 % volume fraction of DMSO leads to the growth of a grain-like structure. It can be concluded that the electrodeposits show worse morphology with $ZnCl_2$ electrolytes as compared to $ZnSO_4$ and $Zn(OTf)_2$ electrolytes. To further investigate the stability of the Zn metal surface in the electrolytes, the Zn metal anodes were soaked into the 2M $Zn(OTf)_2$ electrolyte for a week. XRD and SEM characterizations were performed, as depicted in **Figure S3**. XRD measurements reveal no obvious side reaction for the (101) dominated D-Zn. However, for the commercial Zn metal foil, multiple side peaks are significantly enhanced, corresponding to $Zn_xO_yTf(OH)_{2x-y} \cdot nH_2O$[14]. This is also confirmed by the SEM image, where a terrace-like compact surface can be found for the D-Zn, showing no essential change in the surface morphology after the soak. Whereas for the C-Zn, disordered sub-micron flake-like crystal planes can be discovered after the soak. This shows that the side reactions can occur on the commercialized Zn foil surface, while they are effectively inhibited for the electrodeposited (101)-surface.

*In-situ optical observation*. In-situ optical observation of the electrodeposition and stripping processes is further given in **Figure 2** to study the dynamic plating and striping behavior of the Zn metal on the Cu surface with different electrolytes (e.g., 1.3 M of $ZnCl_2$, 1.3 M $ZnCl_2$ with 18.75% DMSO, 2 M $ZnSO_4$, and 2 M $ZnSO_4$ with 5% DMSO). To slow down the processes for better observation and comparison, the plating and stripping processes were under a relatively low current density of 21 mA cm$^{-2}$. The bare Cu surface is bronze with unidirectional processing scratches under the optical microscope (0.0 s). After electrodeposition for 2.5 s, part of the surface is covered with green mossy Zn metal for all four cases. Interestingly, it can be seen that with a $ZnCl_2$-based electrolyte, the green mossy region is dispersed on the entire electrode surface; whereas with the $ZnSO_4$ electrolyte, the Zn metal tends to nucleate and grow locally. It is also shown that for both $ZnCl_2$ and $ZnSO_4$, adding DMSO leads to a more uniform surface. Consequently, with a longer electrodeposition time (e.g., 100 s), the aggregation of Zn metal can be observed for the cases without DMSO addition. Meanwhile, as a comparison, with DMSO as additives, the Zn aggregation is less severe for both cases. For the electrodeposition using the $ZnSO_4$ electrolyte with DMSO addition, after plating for 100 s, the surface is much flatter than all the other three cases, confirming the previous SEM observation with the formation of a flat and compact surface as compared with other surfaces in **Figure S3**. It is further indicated that the 2 M $ZnSO_4$ electrolyte with DMSO addition is the best choice for the electrodeposition of a thin, flat Zn metal anode. For the stripping process, it is discovered that the Zn metal can be reversibly stripped from the electrode for all four cases after 200 s. Interestingly, as shown in the dotted box, severe pitting corrosion can be observed for the Cu metal surface with pure $ZnCl_2$ electrolyte at the stripping stage, while the addition of DMSO could, to some extent, reduce the Cu surface corrosion. This further suggests that one of the failure mechanisms for the Zn metal battery with $ZnCl_2$ based electrolyte is the corrosion of the Cu current collectors, proving that adding DMSO could help elongate the battery cycle life with ZnCl₂ electrolyte by reducing the Cu surface corrosion.

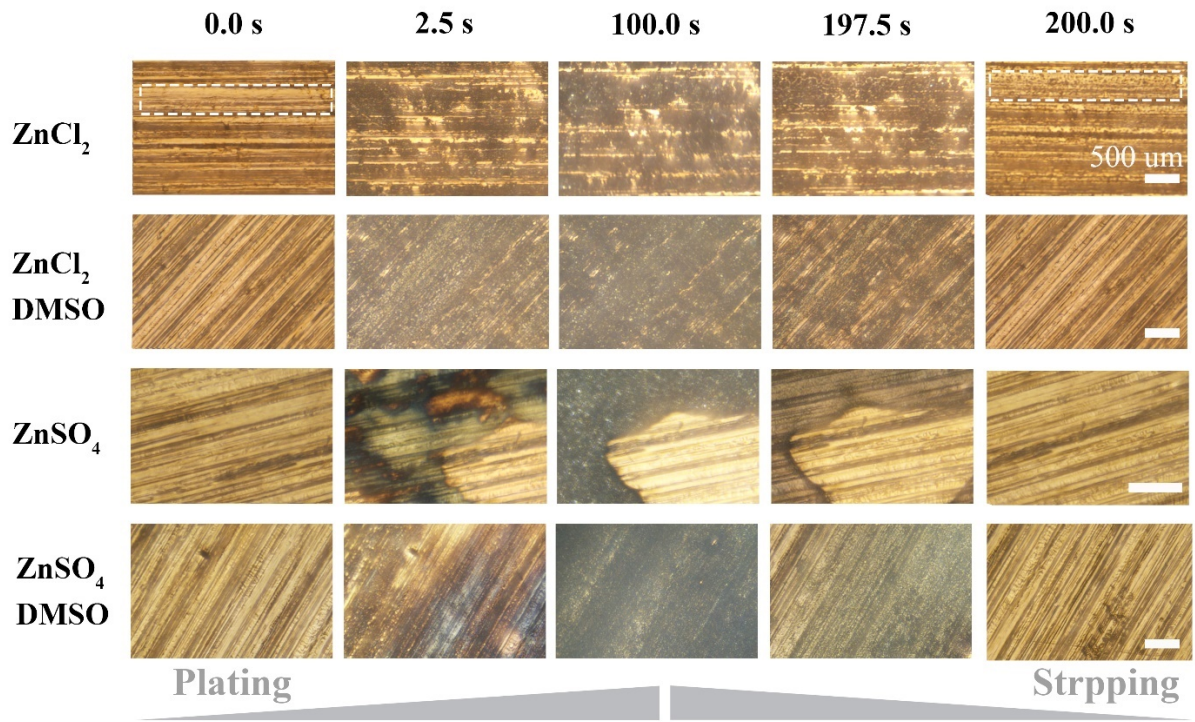

**Figure 2 | Optical microscopic images showing the evolution of the Zn electrodeposits and electrostrips in ZnCl₂ solution and ZnSO₄, with/without DMSO addition**.

*Theoretical understanding by DFT.* To better understand the thermodynamics of the surface/interface formation, the surface and interface energies were calculated using Density Functional Theory (DFT, details in methods). As shown in **Table 1**, the Zn (002) surface has much lower surface energy ($0.016$ eV/Å$^2$) as compared to the Zn (101) surface ($0.037$ eV/Å$^2$), consistent with previous literature[17]. The Cu (111) surface shows slightly lower surface energy as compared to the Cu (001) surface. The four interfacial energies for the Zn//Cu interfaces, namely Zn(002)//Cu(001), Zn(101)//Cu(001), Zn(002)//Cu(111), Zn(101)//Cu(111), are further calculated,. It is shown that the Zn(101)//Cu(001) interface shows a lower energy ($0.024$ eV/A$^2$) as compared to the Zn(002)//Cu(001) interface ($0.032$ eV/A$^2$); whereas the Zn(101)//Cu(111) demonstrates 4 times higher energy with respect to the Zn(002)//Cu(111).

The effect of DMSO addition on the surface population is further investigated through the calculation of the adsorption of DMSO on the two Zn surfaces, as shown in **Figure 3**. It can be observed that for the relaxed configuration, no strong bonding between Zn and DMSO can be discovered for both (001) and (101) surfaces. While the magnified view of the DMSO//Zn (101) and DMSO//Zn (001) surfaces are shown in **Figure 3(c)-3(d)**, respectively. Large surface reconstruction is demonstrated for a Zn (101) surface, while the Zn (002) surface is very stable with the presence of surface DMSO. As shown in **Table 2**, a huge energy reduction (adsorption energy) for the Zn (101) surface can be seen with DMSO (-1.077 eV) as compared to a Zn (002) surface (-0.122 eV). Meanwhile, it is also interesting to note that with pure water on the Zn surface, the adsorption energy is much lower for the Zn (101) surface (-1.157 eV) with respect to the Zn (002) surface, suggesting that water can also leads to the reconstruction of the Zn (101) surface. Moreover, the DMSO+$H_2O$ on a Zn (101) surface shows much lower adsorption energy (-1.692 eV), indicating the synergic effects of DMSO and water on the Zn (101) surface reconstruction.

This study demonstrates that while Zn (002) surface is thermodynamically more stable as compared to the Zn (101) surface, there are two main mechanisms likely responsible for the formation of highly (101)-oriented Zn crystals: (1) When growing on the Cu surface, it has lower interfacial energy with the Cu (111) surface, which could be a preference sites for the nucleation of the Zn (101) surface. Engineering the Cu surface orientation could be an effective method to tune the Zn (002)/ Zn (101) ratio. (2) The inherently unstable characteristic of the Zn (101) surface, which reconstruct with the surface adsorbates ($H_2O$, DMSO), could lead to a higher energy reduction of the system, providing another avenue for the nucleation and growth of the (101) surface, which could explain the increase of (101) plane ratio with the addition of DMSO in the electrolytes as well as the increasing of (101) plane ratio during cycling for the (001)-dominated C-Zn samples (**Figure S4**).

**Table 1| Surface/interface energies for the Zn, Cu surfaces and Zn//Cu interfaces with different crystal orientations.**

| Surface | Zn(002) | Zn(101) | Cu(001) | Cu(111) | Zn(002)// Cu(001) | Zn(101)// Cu(001) | Zn(002)// Cu(111) | Zn(101)// Cu(111) |
|---------|---------|---------|---------|---------|-------------------|-------------------|-------------------|-------------------|
| energy (eV/A$^2$) | 0.016 | 0.037 | 0.092 | 0.080 | 0.032 | 0.024 | 0.011 | 0.047 |

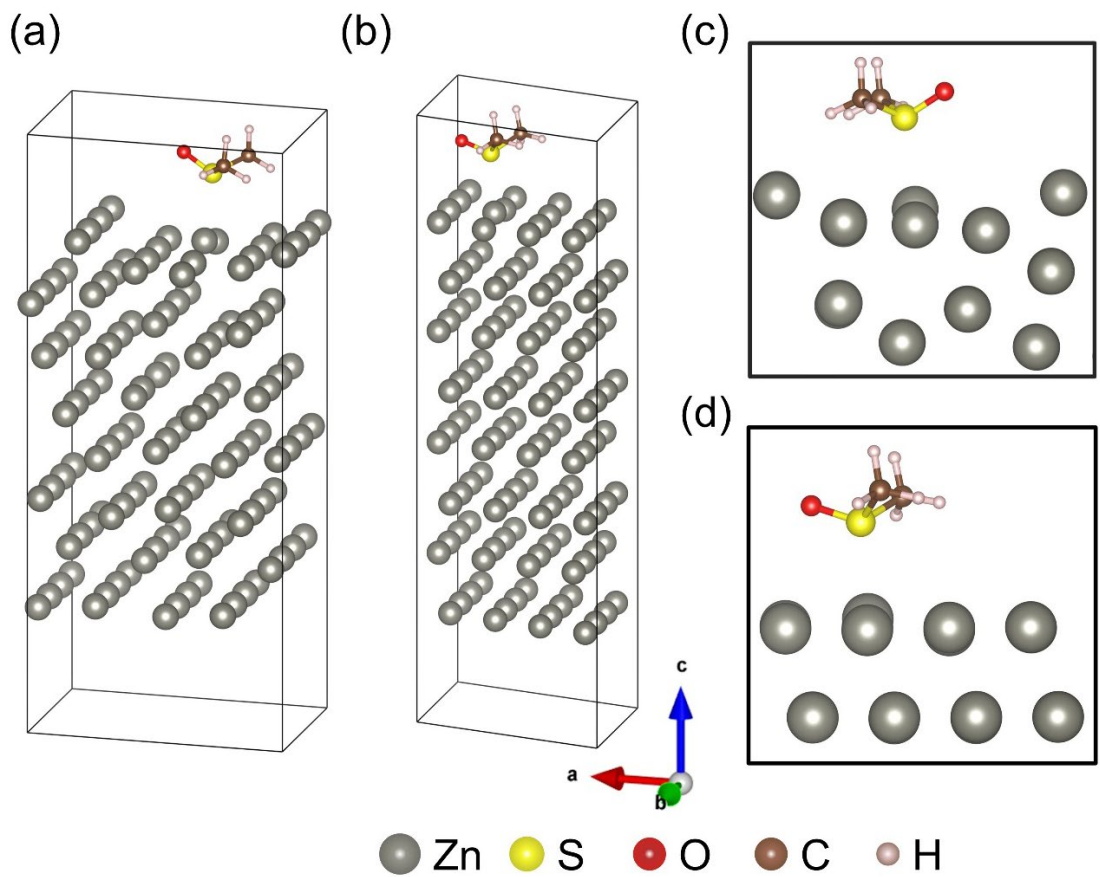

**Figure 3| The relaxed surface structure for DMSO absorbed on the two Zn surfaces as calculated from DFT.** (a) DMSO on Zn (101) surface, the sulfur atom tend to bond with the surface Zn. (b) DMSO on top of a Zn (001) surface, the distance between the Zn and S atoms is larger than the (101) surface. (c) Magnified view for DMSO on Zn (101) surface, showing the large reconstruction of the surface atoms. (d) Corresponding magnified view for DMSO on Zn (001) surface, the surface atoms are similar to those in the bulk.

**Table 2| Calculated absorption energy of the molecules on the different Zn metal surfaces.**

| Configuration | $\Delta E_{abs}$ on (001) surface (eV) | $\Delta E_{abs}$ on (101) surface (eV) |
|---|---|---|
| 3 $H_2O$ | -0.257 | -1.157 |
| 1 DMSO | -0.122 | -1.077 |
| 1 DMSO+1$H_2O$ | -0.319 | -1.692 |

*Surface SEI Composition Characterization*. After understanding the surface morphology and dynamic evolution during electrodeposition, the surface composition (e.g., possible surface SEI composition) is investigated using Raman, XPS, and ToF-SIMS. As shown in **Figure 4(a)**, the Raman spectra for the four as-deposited Zn metal anode indicates that adding DMSO in $ZnCl_2$ could facilitate the formation of ZnO, which exhibits a significant increment of the modes with a Raman shift between 500 cm$^{-1}$ to 600 cm$^{-1}$, corresponding to the $E_2$ and $A_1$ (LO) mode[15]. Meanwhile, no apparent peaks can be observed between 200 cm$^{-1}$ to 1100 cm$^{-1}$ for the three other cases, especially in the case of adding DMSO in $ZnSO_4$. This is further confirmed by the XPS characterization (**Figure 4b**). The peak at ~530.5 eV is attributed to the Zn-O bond[16]. It can be observed that adding DMSO to the $ZnSO_4$ electrolyte could reduce the tendency to form a Zn-O bond, as compared to the case with pure $ZnSO_4$ electrolyte. After Argon etching for 120 s, the Zn-O bonding fraction increases to 50 % with pure $ZnSO_4$ electrolyte, while with DMSO addition, the Zn-O bonding fraction only has 28 %. Meanwhile, an opposite trend can be discovered with the $ZnCl_2$ electrolyte, where adding DMSO could facilitate the Zn-O bonding. To further characterize the possible surface SEI composition, time-of-flight secondary ion mass spectrometry (ToF-SIMS) was performed. As shown in **Figure 4(c)**, three main surface species can be identified, namely $O^{2-}$, $OH^-$, and $S^{2-}$. The obvious enhancement in the S-based components can be observed with $ZnSO_4$-based electrolytes as compared to the $ZnCl_2$-based electrolytes. Meanwhile, notably, with the addition of DMSO in a $ZnSO_4$ electrolyte, the O and S peak intensity decreases, showing that adding DMSO reduces the amount of $OH^-$ and $S^{2-}$ in the SEI of a Zn metal anode.

In a short conclusion, from the three measurements of the as-deposited Zn metal, adding DMSO in $ZnCl_2$ could facilitate the formation of ZnO-based SEI on the metal surface, while $OH^-$ and S-based SEI can be obtained with DMSO in $ZnSO_4$. These surface SEI components could protect the Zn metal surface from corrosion and mechanically suppress the dendrite growth during electrodeposition.

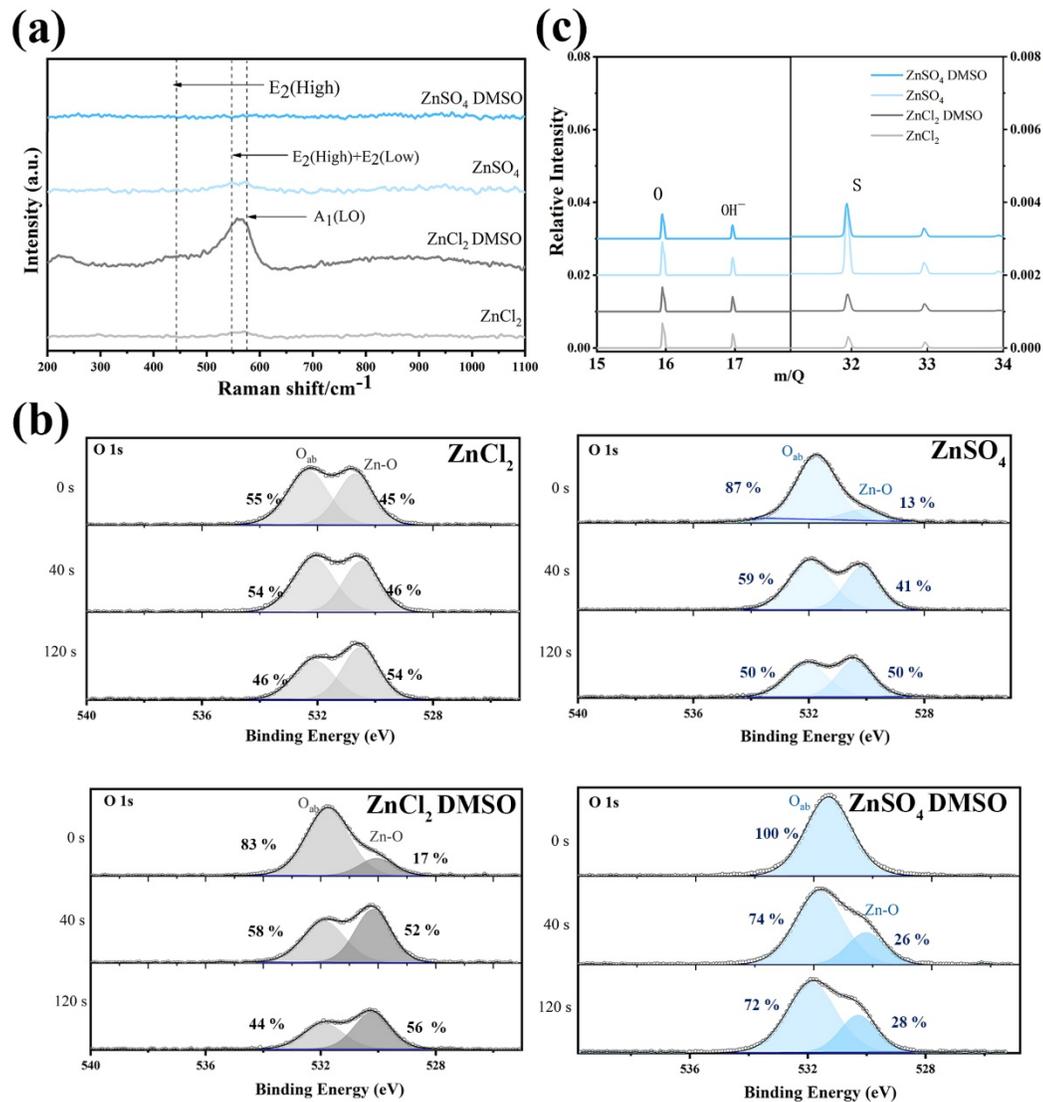

**Figure 4 | Surface composition characterization of the deposited Zn film.** (a) Raman characterization of four electrodes. (b)XPS characterization of the absorbed O and ZnO on four electrodes. (c)ToF-SIMS characterization of the O and S elements of the four electrodes.

*Wettability Analysis*. To further investigate the electrolyte/metal anode surface, the wettability of the electrolytes on the metal surfaces was further studied with the optical contact angle analysis (**Figure 5**). A sessile drop was utilized to quantify the wettability between metals and solutions. Typically, a smaller contact angle reflects better surface wettability and lower surface tension. It can be observed that for the ZnCl₂, ZnSO₄ with DMSO, and Zn(OTf)₂ electrolytes, the D-Zn shows a lower contact angle as compared to the C-Zn, indicating better wettability for the D-Zn with these electrolytes. In contrast, an opposite trend is discovered for $ZnCl_2$ with DMSO, $ZnSO_4$, and $Zn(OTf)_2$ with DMSO. In particular, the contact angle for D-Zn and $ZnSO_4$ with DMSO exhibits the lowest contact angle of ~68°. The smaller contact angle for D-Zn anodes with the $ZnSO_4$ containing DMSO and $Zn(OTf)_2$ electrolytes indicates that they have better wettability with the Zn surface, which could lower the nucleation barrier for the Zn electrodeposition during the cycling process. This study suggests that DMSO could be an effective electrolyte additive for $ZnSO_4$-based and $ZnCl_2$-based electrolytes. Meanwhile, an increase in the contact angle from 84° to 90° is observed for C-Zn anode with DMSO addition in $Zn(OTf)_2$ electrolyte. Moreover, an even larger increase of the contact angle is shown for D-Zn anode with DMSO addition in $Zn(OTf)_2$ electrolyte (from 78° to 108°).

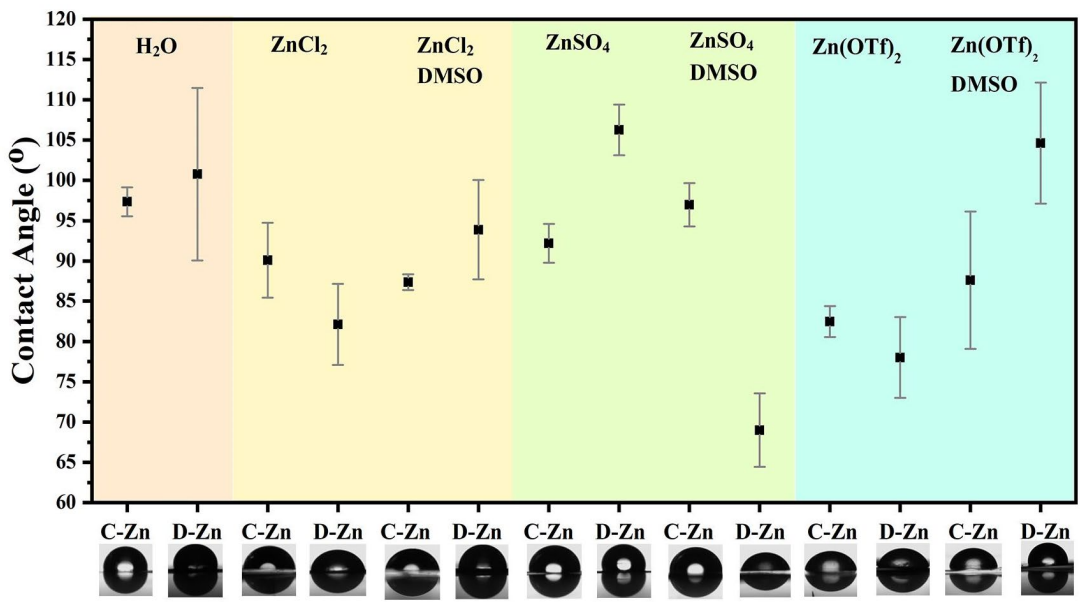

**Figure 5| The statistical data and images of the optical contact angles.** The different electrolytes dropped on commercial Zinc foils (C-Zn) and Zn deposited on copper foils (D-Zn) were shown in the graph.

*Electrochemical performances.* The Electrochemical performances of the as-deposited thin electrodes are examed. A half cell was assembled, with a 5 μm D-Zn electrode against the 20 μm C-Zn electrode. The charge and discharge depth are 1 μm (20% DOD) and the current density is 0.5 mA cm$^{-2}$. A higher discharge current (5 mA cm$^{-2}$) was applied to form enough nucleation sites for the first 10 cycles. As illustrated in **Figure 6 (a)**, the (101) D-Zn exhibits a low overpotential (< 50 mV) for 200 h. In contrast, the (002) D-Zn cell appears a typical dendrite growth overpotential from 70 h, which transforms into a soft short circuit voltage profile since 82 h. To get insight into the better cycling performance of the asymmetric cells. The contact between the current collector and the deposition layer as well as the surface condition was further studied by the electrochemical impedance spectrum. For a C-Zn‖C-Zn cell, one semi ellipse can be found in the spectrum (**Figure 6b**) and the resistance is 601 $\Omega$ cm$^{-2}$ because no ascensional current collector was used. While two semi ellipses appear for the D-Zn‖C-Zn cells. The first semi ellipse that appears in high frequency (> 1000 Hz) represents the charge transfer impedance of the electrodeposition interface and the second one in lower frequency is the electrode surface. For both surfaces, the (101) D-Zn shows smaller charge transfer resistance ($R_{ei}$ = 44 $\Omega$ cm$^{-2}$, $R_{sf}$ = 283 $\Omega$ cm$^{-2}$) as compared to the (002) D-Zn ($R_{ei}$ = 53 $\Omega$ cm$^{-2}$, $R_{sf}$ = 371 $\Omega$ cm$^{-2}$), suggesting that the (101) D-Zn electrode has better electron connection and ion transfer in the two interfaces. To assess the nucleation barrier on the surface during the cycling, a 10 mA cm$^{-2}$ current was used for the galvanostatic test. As shown in **Figure 6(c)**, the voltage of the (101) D-Zn reached the plateau quickly, and all these three electrodes reached the typical Zn metal growth stage in 200 s. The overpotential of the (101) D-Zn is 75 mV, which is smaller than the (002) D-Zn (86 mV) and commercial zinc foil (95 mV). The result is consistent with the conclusion of optical contact angle analysis. The V$_2$O$_5$‖Zn cells were further fabricated to explore the performance of the thin anode in a full cell. As shown in **Figure 6(d)**, the specific capacities increase in the first 100 cycles for all the cells during the V$_2$O$_5$ activation process, in good agreement with the previous reports [18]. The 5 μm (101) D-Zn exhibits a similar curvilinear trend with 20 μm C-Zn, but with a slightly steep slope after reaching the peaks. The cell assembled with a 5 μm (101) D-Zn anode maintains good capacity without nonlinear decay in 200 cycles, whereas the 5 μm (002) D-Zn cell shows a sudden deterioration in the 179[th] cycle. For the cells assembled with 3 μm and 2 μm D-Zn, the nonlinear decay occurs in the 146[th] and 114[th] cycles. In contrast, the failure processes are earlier in (002) D-Zn. Details of capacities change are illustrated in **Figure 6(e)-6(g)**. The disappearance of the second discharge platform from 0.67 V to 0.45 V, which is highly related to the Zn$^{2+}$ insertion process[19], can be observed in the deterioration stage of a (002) D-Zn‖V$_2$O$_5$. That can be ascribed to the loss of Zn-ion and the poor reversibility on the surface. By contrast, the (101) D-Zn‖V$_2$O$_5$ maintains the same charge and discharge voltage property with a small change of polarization.

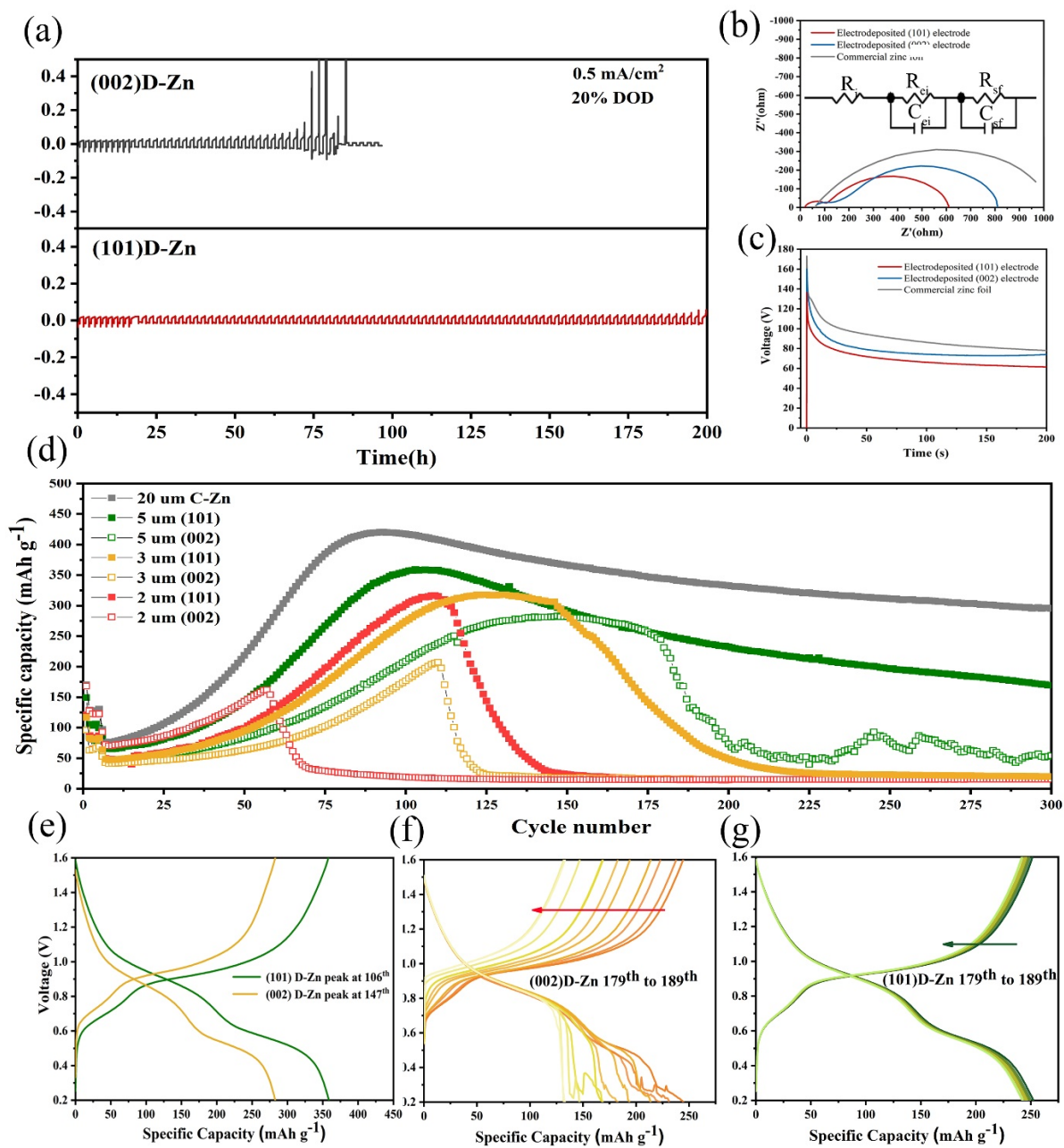

**Figure 6| Electrochemical performances for the as-deposited thin Zn metal anode.** a) Voltage profile of 5 μm Zinc electrode against 20 μm Zn metal electrode asymmetric cell cycling under 0.5 mA cm$^{-2}$. b) EIS results of electrodeposited electrodes and pure Zn metal foil against Zn metal foil asymmetric cell. c) Nucleation overpotential results of asymmetric cells with electrodeposited (101), (002) and commercial zinc foil electrodes against Zn metal foil at 10 mA cm$^{-2}$. d) Cyclic capacity of Zn‖V$_2$O$_5$ full cells . Charge and discharge curves of 5 μm D-Zn‖V$_2$O$_5$ cell  with e) highest capacity, and from 179th to 189th cyclic of f) (002) D-Zn, and g) (101)D-Zn.

**Conclusion**

In conclusion, the surface orientation of the electrodeposited thin Zn metal anode (down to 2 μm) can be controlled by the electrolyte additive, where the (101)-oriented surface dominates with the addition of DMSO molecule as compared to the (002)-oriented surface without electrolyte additives in $ZnSO_4$ aqueous solution. SEM observation indicates the formation of a flat terrace-like compact surface for the (101)-oriented surface. Insitu optical observation confirms the electrodeposition of a compact surface with the addition of DMSO, which can be reversibly stripped. Meanwhile, surface corrosion on the Cu metal current collector surface is observed with the $ZnCl_2$ electrolyte. DFT calculations indicate that although the Zn (101) surface has higher surface energy, a large reconstruction is observed for the Zn (101) surface with DMSO and $H_2O$ absorption, which could largely lower the Zn (101) surface energy and nucleation barrier for the Zn (101) surface. Raman, XPS, and ToF-SIMS characterizations indicate that adding DMSO in $ZnCl_2$ could facilitate the formation of ZnO-based SEI on Zn metal surface, while $OH^-$ and S-based SEI can be obtained with DMSO in $ZnSO_4$. The electrochemical testings were performed, which demonstrates a higher cyclability for the (101) oriented Zn in the half cell for the same electrode thickness and lower charge transfer barrier. $Zn\|V_2O_5$ full cells were further assembled, showing better capacity retention for the (101)-Zn as compared to the (002)-Zn with the same thickness (5 μm, 3 μm, and 2 μm). This study paves the way towards the application of thin Zn metal anode in Zn metal batteries.

4. Methods

*Preparation of ultra-thin Zinc electrodes*. The ultra-thin zinc electrodes were prepared through galvanostatic electrodeposition using an Iviumnstat workstation. The electrodeposition parameters conformed to faradays law. The experiments were performed under a two-electrode system, including a Zinc plate (99.995%) as the working electrode and a copper foil (99.99% with electropolishing) as the counter electrode. The electropolishing process was conducted with a two-electrode system. Both electrodes were copper foils, using 55% $H_3PO_4$ (diluted with >85.0% from Sinopharm Chemical Reagent Co.) aqueous solution as electrolyte.

We prepared the electrodeposition solutions with 1) 2M $ZnSO_4$(≥99.995%, from Aladdin Bio-Chem Technology Co.)with/without 5vol% the dimethyl sulfoxide (DMSO) (>99.5% from Sinopharm Chemical Reagent Co.), 2) 1.3 M $ZnCl_2$ (≥ 99.995% from Aladdin Bio-Chem

Technology Co) with/without 18.75vol% DMSO and 3) 2M $Zn(OTf)_2$ (98%, Sigma-Aldrich Macklin Biochemical Co.) with/without 5vol% DMSO.

*Characterizations.* The morphologies of the copper foils and electrodeposited zinc electrodes were detected by SEM (Hitachi S4800FESEM)equipped with EDX (EDAX) and SPM (Veeco diInnova). For cross-sectional SEM images, electrodeposited zinc electrodes were cut by surgical knife along with mechanical scratches, and Argon ion milled by a JIT CP-200E. The XRD data were collected to detect the intensity of crystal planes and by-products on a SmartLab using Cu Kα radiation (λ=1.54059 Å) with a scan rate of 10°/min. The Raman spectrum was recorded by Renishaw in the 200~1100 cm-1 using a 532 nm laser with 1200-line mm$^{-1}$ grating. The power of the laser was set to 10 mW to intensify the signals. The XPS measurements to obtain the species of SEI were carried out on a Thermo Scientific K-Alpha with a monochromatic Al K$\alpha$ ($hv$ = 1486.6 eV) excitation source. The spectrum was analyzed using CASA XPS. The mass spectra were collected using time-of-flight secondary ion mass spectrometry (ION TOF ToF SIMS 5-100). The samples were rinsed with DI water and ethanol thoroughly for pretreatment and then dried at room temperature within 30 min. Contact angles between the electrolytes and commercial/electrodeposited electrodes were measured with Dataphysics OCA 20 applying the sessile drop method and fitted with the ellipse method.

*In-situ Optical Microscope Observation.* The in-situ optical virtualization observations are recorded on an optical microscope (Sunny Optical Technology) assisted with CHI-instrument electrochemical workstation. The in situ electrodepositing and electrical stripping processes were conducted in a homemade electrolytic bath, and the top view of an electrode interface is obtained.

*Electrochemical Test.* Zn‖Zn asymmetric cells were assembled with 2 μm‖20 μm, 3 μm‖20 μm, and 5 μm‖20 μm zinc electrodes sandwiching the 2 M $Zn(OTf)_2$ (98%, Sigma-Aldrich) as the electrolyte and filters (Mixed cellulose ester membrane, Shanghai Xingya purification material factory ) as the separators in CR2025-type coin cells. The 2 μm/3 μm/5 μm zinc electrodes were pre-deposited while the 20-μm-thick electrodes were commercial foils (99.995%). For full cell test, $V_2O_5$ without further treatments  (99%, Macklin Biochemical Co.) was mixed with Ketjen black and CMC (Carboxymethyl Cellulose) in a weight ratio of 7:2:1 using water as solvent. The slurry was pasted on a carbon fiber cloth (HCP010N, Shanghai Hesen Electric Co., Ltd) and air-dried at 60 ℃ for 12h. The full cells were discharged and charged at 80 mA g$^{-1}$ for the initial 5 cycles and 400 mA g$^{-1}$ for the rest cycles. The asymmetrical-cell cycling tests were conducted by using 5 μm electrodeposited (101) and (002) electrodes against 20 μm commercial zinc foils. The cycling strategy applies a 5 mA cm$^{-2}$ discharge current to reach the depth of 0.5 μm (0.293 mAh cm$^{-2}$), followed by another 0.5 μm under 0.5 mA cm$^{-2}$, and then charged to 1 μm (0.586 mAh cm$^{-2}$) with 0.5 mA cm$^{-2}$. For the rest of the cycling process, a constant current of 0.5 mA cm$^{-2}$ is set for charge and discharge[20] . The nucleation overpotential was obtained by electrodepositing on three sorts of electrodes in the galvanostatic method at 10 mA cm$^{-2}$ for 200 s. The charge-discharge experiments were performed on a Land CT3001A battery test system at 25℃.

*DFT Calculations.*

In this study, all DFT calculations were conducted via open-source GPAW software[21], using the projector-augmented wave (PAW)[22] method in a finite-difference (FD) way. We use the PBE exchange-correlation functional[23] and Monkhorst-Pack scheme for the surface, interface, and adsorption calculations. For surface calculations, the 4-layer (002) Zn slab, (101) Zn slab, (001) Cu slab, and (111) Cu slab were built with 96 atoms, 48 atoms, 80 atoms, and 64 atoms. The k-points meshes were respectively chosen as $1×3×1$, $2×3×1$, $1×3×1$, and $1×3×1$. As for interface calculations, all interfaces were constructed by 4-layer Zn slab and 4-layer Cu slab with less than 10% mismatch. In details, the interfaces area and *k*-point meshes are set with (18.16×7.56 Å$^2$, $1×3×1$) for Zn(002)//Cu(001), (7.56×11.15 Å$^2$, $3×2×1$) for Zn(101)//Cu(001), (5.19×8.99Å$^2$, $5×3×1$) for Zn(002)//Cu(111), and (10.84×8.38 Å$^2$, $2×3×1$) for Zn(101)//Cu(111). Finally, to verify the reconstruction effect introduced by DMSO and water, we used 8-layer (002) Zn slab and (101) Zn slab with both 128 atoms. Before adding DMSO and water, the (002) Zn slab and (101) Zn slab was fully relaxed without large reconstruction. The cutoff force for all DFT calculations was set to below 0.05 eV A$^{-1}$.

**Supporting Information**

Supporting Information is available from the Wiley Online Library or from the author.

**Conflict of Interest**

The authors declare no conflict of interest.

**Acknowledgements**

This work was financially supported by the Fundamental Research Funds for the Central Universities (2020FZZX003-04, 2021FZZX003-02-03, Z. H.). A startup grant from Zhejiang University is acknowledged (Z. H.).

Received: ((will be filled in by the editorial staff))

Revised: ((will be filled in by the editorial staff))

Published online: ((will be filled in by the editorial staff))

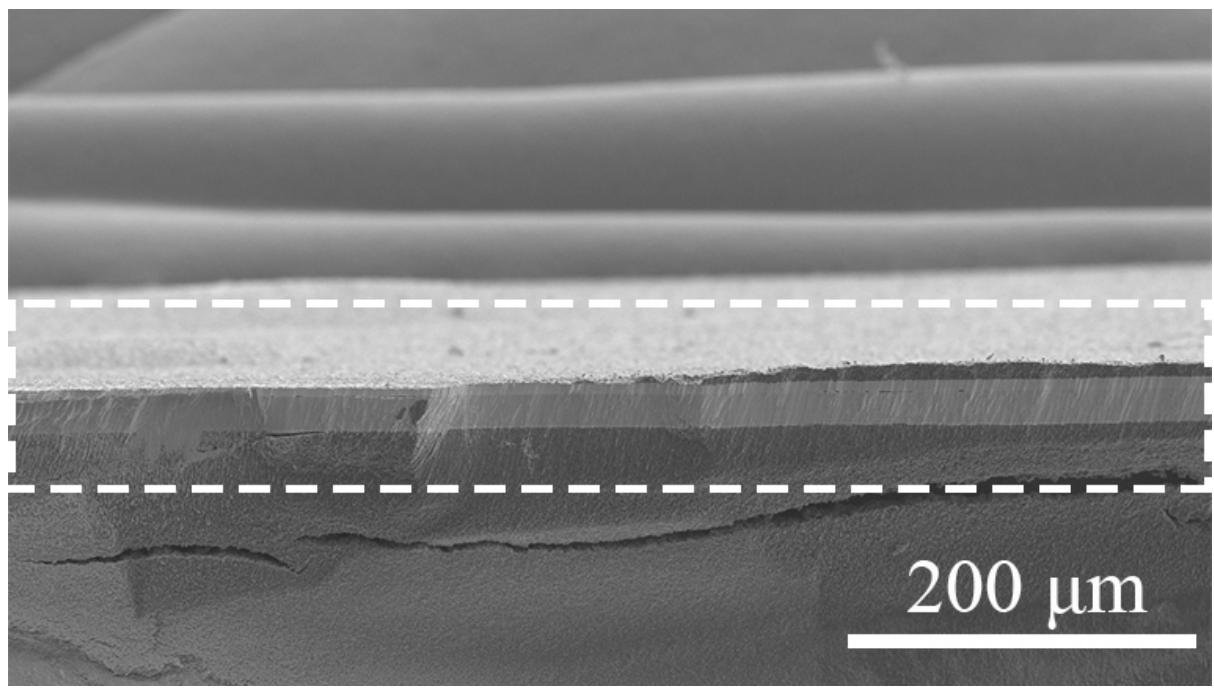

**Figure S1.** SEM images of ultra-thin electrode in cross-section view.

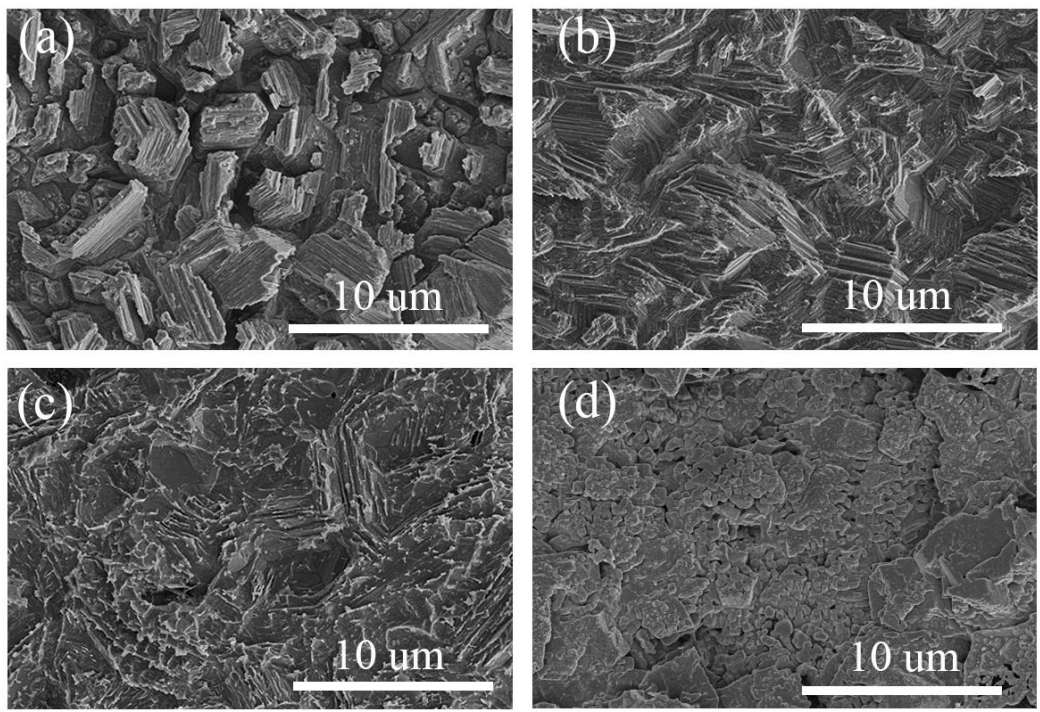

**Figure S2**. SEM images illustrate the morphorlogies of Zn electrodoposits in (a) 1.3 M $ZnCl_2$ solution, (b) 1.3 M $ZnCl_2$ with 18.75% DMSO, (c) 2M $Zn(OTf)_2$ solution, (d)2M $Zn(OTf)_2$ with 5% DMSO.

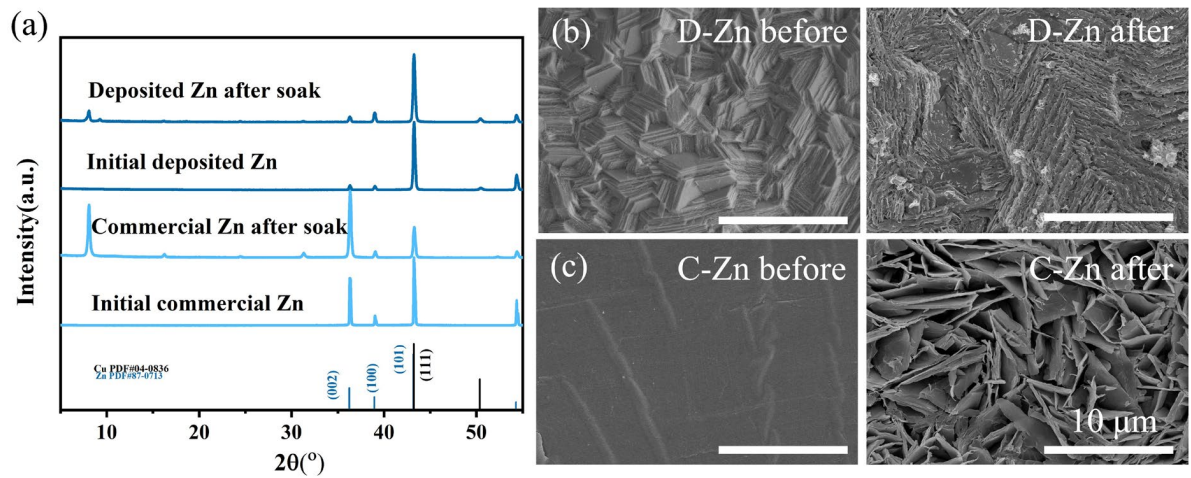

**Figure S3** (a) XRD patterns of C-Zn and D-Zn before and after soak in the Zn(OTf)₂ electrolyte. (b) SEM images of D-Zn before and after soak in the 2M Zn(OTf)₂ for a week. (c)SEM images of C-Zn before and after soak in the 2M Zn(OTf)₂ for a week.

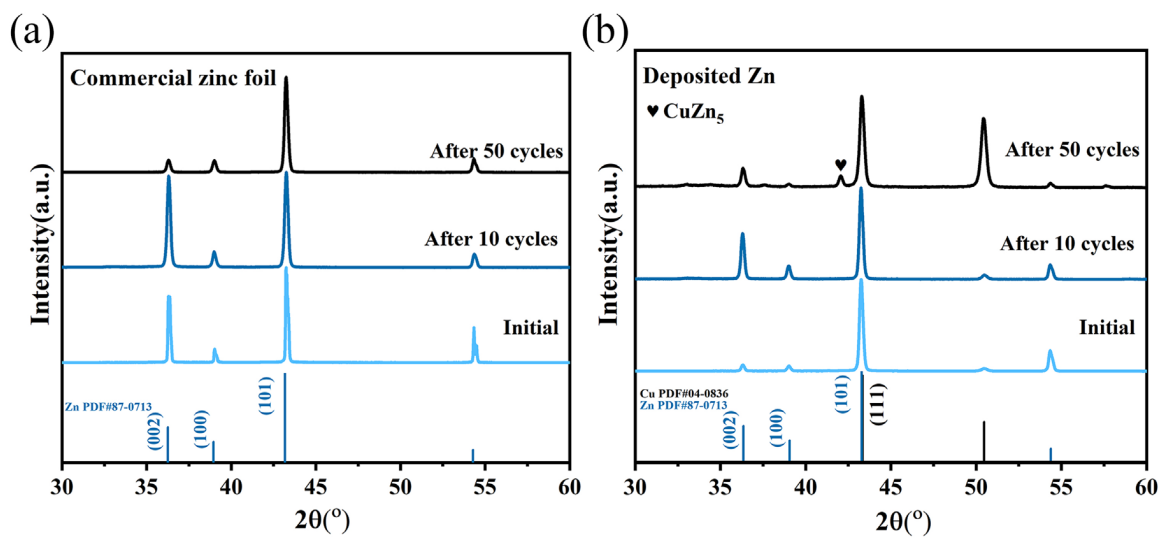

**Figure S4** XRD patern of C-Zn and (101) D-Zn before and after cycling. (a) XRD paterns of initial C-Zn and electrodes after 10 and 50 cycles. (b) XRD paterns of initial D-Zn and electrodes after 10 and 50 cycles.